# Deep Learning Algorithm for Advanced Level-3 Inverse-Modeling of Silicon-Carbide Power MOSFET Devices

**Massimo Orazio Spata[1], Sebastiano Battiato[1], Alessandro Ortis[1], Francesco Rundo[2], Michele Calabretta[2], Carmelo Pino[2], Angelo Messina[2]**

[1]Dipartimento di Matematica ed Informatica, University of Catania
[2]STMicroelectronics ADG R&D Division Catania, Italy

massimo.spata@unict.it, sebastiano.battiato@unict.it, ortis@dmi.unict.it
francesco.rundo@st.com, michele.calabretta@st.com, carmelo.pino@st.com,
angelo.messina@st.com

**Abstract.** Inverse modelling with deep learning algorithms involves training deep architecture to predict device's parameters from its static behaviour. Inverse device modelling is suitable to reconstruct drifted physical parameters of devices temporally degraded or to retrieve physical configuration. There are many variables that can influence the performance of an inverse modelling method. In this work the authors propose a deep learning method trained for retrieving physical parameters of Level-3 model of Power Silicon-Carbide MOSFET (SiC Power MOS). The SiC devices are used in applications where classical silicon devices failed due to high-temperature or high switching capability. The key application of SiC power devices is in the automotive field (i.e. in the field of electrical vehicles). Due to physiological degradation or high-stressing environment, SiC Power MOS shows a significant drift of physical parameters which can be monitored by using inverse modelling. The aim of this work is to provide a possible deep learning-based solution for retrieving physical parameters of the SiC Power MOSFET. Preliminary results based on the retrieving of channel length of the device are reported. Channel length of power MOSFET is a key parameter involved in the static and dynamic behaviour of the device. The experimental results reported in this work confirmed the effectiveness of a multi-layer perceptron designed to retrieve this parameter.

## 1. Introduction

Inverse modelling is a process which allows to retrieve the underlying model parameters of a physical system starting from its behavior observations. Related to a power MOSFET (metal-oxide-semiconductor field-effect transistor) [10], inverse modelling leverages deep learning algorithms to predict the device's characteristics from input features embedded in the input device signal. Related modelling for a power MOS is important for several reasons. Firstly, accurate models help to determine the device performance under various operating conditions and optimize the design. Secondly, models allow simulation of the MOS behavior in the overall circuit, enabling the prediction of the circuit performance and its optimization. Finally, it is possible to enable the assessment of the thermal and electrical stresses on the device and aid in reliability analysis.

Deep learning technology is helpful for power MOS modelling because it allows complexity handling, whereas traditional models may not be sufficient to capture the complex behavior of modern power

MOS devices. Deep learning algorithms can handle the intrinsic complexity providing more accurate modelling. They also allow large data handling. Indeed, deep learning algorithms can handle large amounts of data, which is crucial for modelling complex power MOS devices that have many parameters and operational modes. It also allows automation: deep learning algorithms can automate the modelling process, reducing the time and resources required for manual modelling. Finally deep learning algorithms can model nonlinear behavior of power MOS devices, which traditional models may not be able to capture accurately [11]. Furthermore, such models aid in the development of new fabrication processes and their optimization. Specifically, this work is focused on the analysis of the Silicon Carbide Power MOSFET (SiC Power MOS).

SiC (Silicon Carbide) power MOSFETs (Metal-Oxide-Semiconductor Field-Effect Transistor) are a type of power electronic device that are used for high-power and high-temperature applications. They have several key characteristics that make them different from traditional silicon MOSFETs: high breakdown voltage: SiC MOSFETs can handle much higher voltages than silicon MOSFETs, making them useful in high voltage applications. High temperature operation: SiC MOSFETs can operate at much higher temperatures than silicon MOSFETs, making them suitable for high-temperature environments. High switching speed: SiC MOSFETs have a faster switching speed than silicon MOSFETs, which can improve the overall efficiency of power electronic systems. High thermal conductivity: SiC MOSFETs have a higher thermal conductivity than silicon MOSFETs, which allows them to dissipate heat more effectively and improve their overall reliability. Low gate charge: SiC MOSFETs have a lower gate charge than silicon MOSFETs, which reduces the amount of energy required to switch them on and off. Low on-resistance: SiC MOSFETs have a lower on-resistance than silicon MOSFETs, which increases the efficiency of the power conversion. Robustness: SiC MOSFETs are more robust to over-voltage and over-current conditions than Silicon MOSFETs.

Overall, SiC power MOSFETs are designed for high power, high temperature, and high frequency switching applications, making them suitable for use in power electronics systems such as inverters, converters, and motor drives [2]. There are different methods to be used to perform inverse modelling of a power MOSFET using deep learning algorithms. Such methods are based on the usage of such supervised approaches [11]. In order to create a SiC power MOSFET model has been used Matlab Simulink. Simulink device modelling consists in computer-aided device simulations able to capture the physical and electrical behavior of semiconductor devices, helping semiconductor industry to drastically reduce prototyping costs while achieving desirable device properties. Due to the increasing device design complexity, simulations help engineers to create a better device with specific electrical properties according to device and materials standards.

More specifically, through the usage of the sampled static behavior of the MOSFET, the designed deep architecture has been trained to retrieve physical parameters, in this case, the actual channel length Lp of the device. More in detail, we collect the SiC Power MOSFET dynamic behavior related to drain current versus the drain-source voltage, according to the polarization voltage (gate-source voltages). This set of signals has been used as input dataset. The output dataset to be learned is channel-length of the analyzed device. To generalize the correlation between input data and channel length of the device, we have designed a deep backbone based on the usage of Multi-Layer Perceptron (MLP) [12]. The remainder of this paper is organized as follows: Section II provides a summary of previous research works related to the topic, the proposed pipeline which and the details of the developed deep learning model architecture as well as a brief description of a power MOS model is presented in Section III. Section IV presents the experimental results, which confirmed the highly promising performance of the designed solution.

## 2. Related work

The existing prior art includes several promising proposals as reported in [2]-[7]. In particular, the models proposed in [2,3,4] achieve a very good fit for the device's static curves and accurately predict its transient behavior. These efforts have also been extended into extensive numerical modelling of SiC MOSFET's [5,6,7]. However, these models are generally complex both in terms of implementation and parameter extraction and often require proprietary software for the parameter's extraction phase.

In [2], modeling of the SiC power MOSFET was analysed with an automated tuning process, developing a MATLAB's script Genetic Algorithm to adjust the values of user-specified model parameters until agreement with characterization data is obtained. A direct comparison of switching, juxtaposed with empirical outcomes derived from double-pulse testing, showcased the adeptness of the formulated model in forecasting key temporal characteristics and switching losses. Of greater significance, the established model displays computational efficiency attributed to the fundamental simplicity of the Level-3 MOSFET model that forms its core. As a result, the fundamental contribution of this research lies in the creation of a modelling approach tailored for SiC power MOSFETs, meticulously fine-tuned for optimizing power electronics application design.

In [3], the authors presented an innovative approach to crafting a behavioral SPICE model for a SiC MOSFET, employing an automated tuning procedure. This technique facilitated the real-time adjustment of both static and dynamic traits, resulting in a favourable concurrence.

The advantages of the suggested strategy are double. Initially, the tuning procedure is autonomously executed, obviating the necessity for the designer to manually fine-tune (and repeatedly adjust) parameters until satisfactory alignment is achieved. Secondly, the devised approach for creating device models is swift and yields precise behavioral models. The static curve fittings achieved in each stage are equivalent to those generated by commercially accessible software platforms, and the alignment of transient data is akin to the meticulous and time-consuming manual interactions often required.

In [4], unlike the conventional approach to developing power FET models, the authors formulated a method that capitalizes on both static and dynamic characterization data pertaining to the subject device. Merging signal processing techniques and derivative-free global optimization methods, it becomes possible to create high-fidelity models that precisely forecast switching behavior, all without requiring the laborious process of manually fine-tuning parameters. The model employed to showcase the fitting procedure is verified using an independent dataset that switching data at a distinct drain current from the one utilized during the tuning phase. The model refined in real-time exhibits enhanced alignment compared to the statically adjusted model for both sets of data. Nevertheless, the aforementioned methods underscore the drawbacks involving additional computational time, thereby diminishing their utility for application designers' perspective. Conversely, this article introduces a streamlined Power MOS Level-3 inverse modelling approach, leveraging a Multi-Layer Perceptron and fine-tuned for power electronics applications, thus acknowledging and addressing the requirements of application designers.

### 3. Proposed pipeline

As previously introduced, we have designed a Multi-Layer Perceptron suitable retrieve the channel length of SiC Power MOSFET starting from its static dynamic based on Level-3 model. Before to describe the proposed system, a brief introduction on Level-3 modelling will be made.

Figure 1 shows the proposed architecture pipeline, which is described in detail in the following paragraphs. A power MOSFET (metal-oxide-semiconductor field-effect transistor) is a type of transistor that is commonly used to control high-power electronic devices such as motors, power supplies, and other types of industrial equipment. The level of a power MOSFET refers to the voltage rating of the device, and is typically divided into three levels: Level 1, Level 2, and Level 3.

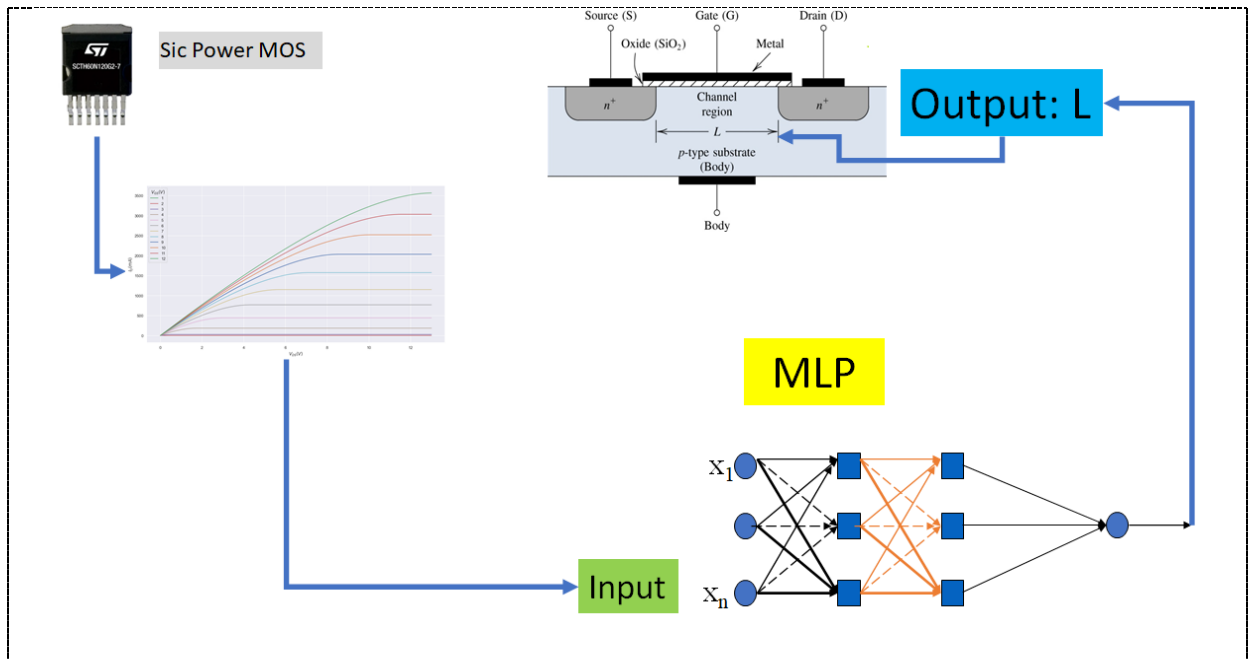

**Figure 1.** SiC Power MOSFET pipeline for the extraction of the physical parameter Lp.

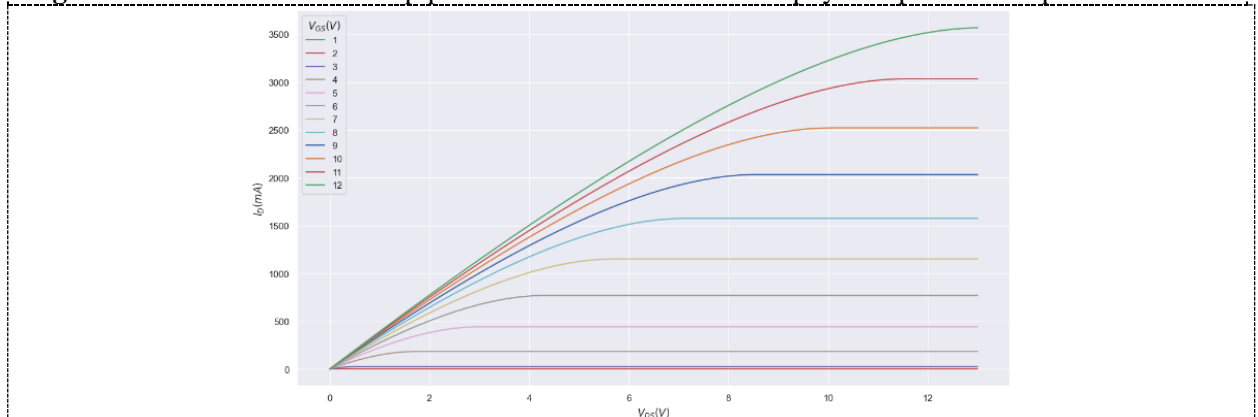

**Figure 2.** SiC Power MOSFET static curves driven by gate-source voltages.

*Level 1* power MOSFETs are typically rated for low voltage applications (typically 600V or less) and can handle relatively low current levels (typically less than 100A).

*Level 2* power MOSFETs are rated for higher voltage applications (typically 600V to 1,200V) and can handle higher current levels (typically up to 200A).

*Level 3* power MOSFETs are rated for the highest voltage applications (typically 1,200V or higher) and can handle very high current levels (typically more than 200A). They are designed for high power applications such as in inverters, welders, and high-power motor drives. The Level 3 model originates from empirical correlations established between practical data acquired through experiments and the pre-existing theoretical models [10]. Close to the Level 2 model structure, the suggested Level 3 model adopts a semi-empirical strategy that prioritizes the process of parameter extraction. Incorporated within are drain-induced barrier lowering (DIBL) and mobility degradation due to lateral field effects [10]. These formulations are relevant to extended-channel devices possessing a gate length around 2 $\mu$m [10]. The fundamental MOS equations [6, 10], linking drain current to drain-source voltage and drain current to gate-source voltage, as explained by equations (1) and (3):

On cutoff region *(V$_{gs}$ < V$_{th}$)*

$$I_{ds} = 0$$

On region (V$_{gs}$ > V$_{th}$)

$$I_{ds} = \begin{cases} \beta(V_{gs} - V_t)V_{ds} - (1 + f_b)\frac{V_{ds}^2}{2} & if\ 0 \leq V_{ds} \leq V_{dssat} \\ \frac{\beta}{2(1+f_b)}(V_{gs} - V_t)^2 & if\ V_{ds} \geq V_{dssat} \end{cases} \quad (1)$$

And:

$$\beta = KP.\frac{W_p}{L_p} \quad (2)$$
$$KP = u_{eff}.COX \quad (3)$$

where:
- $\beta$ describes the temperature dependence of the current-voltage (I-V) characteristics
- *KP* is the intrinsic transconductance parameter
- *COX* is Oxide capacitance per unit gate area.
- *Wp* (or *W*) is the width of the MOSFET
- *Lp* (or *L*) is the channel length of the MOSFET
- *u$_{eff}$* is the mobility of the inversion layer electrons

In Figure 2 an instance of drain-current (*I$_d$*) versus drain-source voltage (*V$_{ds}$*) is reported. Each curve depends of the gate-source polarization of the tested SiC Power MOS. Each curve is known as "transfer-curve" of the Power MOS.

The Power MOSFET transfer curve can be divided into three regions [10]: The saturation region, where the drain-source current (Ids) is almost constant and increases with an increase in the gate-source voltage (*V$_{gs}$*). The linear region, where the drain-source current (*I$_{ds}$*) increases linearly with the gate-source voltage (*V$_{gs}$*). The cutoff region, where the drain- source current (*I$_{ds}$*) is zero, regardless of the gate-source voltage (*V$_{gs}$*). It is generally recommended, for automotive applications, to operate the Power MOSFET in saturation region for best efficiency. To simulate the Level 3 model, we have used a SIMULINK environment provided by MATLAB © framework [12].

*3.1. Level 3 MOSFET's Simulink simulated model*

Starting from mathematical model described in previous paragraph, using Simulink, has been created a dataset with 60000 different transfer-curves of SiC MOSFET, modulating the following parameters of the model (*V$_{th}$*, *L$_p$*, *W$_p$*, *KP*, *R$_d$*, *R$_s$*, $\phi$, $\gamma$, $\theta$, *V$_g$*) as reported in the Eqs (1)-(3). Each parameter can ranging in a specific set as reported in Table 1.

**Table 1.** Model parameters of the simulated Power MOS

| Parameter | Default | Range | | |
|---|---|---|---|---|
| Channel Length (L) | 10$^{-7}$ | 10$^{-7}$ | 5*10$^{-6}$ | m |
| Channel Width (W) | 1 | 10$^{-2}$ | 10 | m |
| Drain resistance (R$_D$) | 10$^{-3}$ | 1*10$^{-4}$ | 10$^{-2}$ | Ohm |
| Source resistance (R$_S$) | 10$^{-3}$ | 1*10$^{-4}$ | 10$^{-2}$ | Ohm |

| Parameter | Default | Range | | |
|---|---|---|---|---|
| Threshold Voltage ($V_T$) | 3 | 2 | 8 | V |
| Transcondutance (K) | $2*10^{-5}$ | $2*10^{-7}$ | 20 | $A/V^2$ |
| Bulk Threshold ($\gamma$) | 0 | 0 | 10 | $V^{0.5}$ |
| Surface potential ($\varphi$) | 0.6 | 0 | 6 | V |
| $V_{GS}$ dependance mobility | 0 | 0 | 10 | 1/V |
| Gate Voltage ($V_{GS}$) | | 1 | 12 | V |
| Temperature | 25 (step=25) | 25 | 175 | Celsius |

By changing the parameters reported in Table I, we have created the dataset of input SiC Power MOSFET transfer curves to be used for the work herein reported (retrieved the channel length of the Power MOS).

*3.2. Multi-Layer Perceptron (MLP)*

3.3. A multi-layer perceptron (MLP) [16] is a type of artificial neural network (ANN) consisting of at least three layers of nodes: an input layer, one or more hidden layers, and an output layer. MLPs are commonly used in supervised learning problems, such as classification or regression tasks, where the network is trained on a set of labeled examples to make predictions on new, unseen data. Each node in an MLP is a mathematical function that takes a set of input values and computes an output value. In the input layer, each node corresponds to a feature or attribute of the input data. In the hidden layers, each node typically applies a nonlinear transformation to the outputs of the nodes in the previous layer. The output layer computes the final output of the network, which can be a single value for regression or a set of values representing the probabilities of different classes for classification. During training, the weights of the connections between the nodes are adjusted using an optimization algorithm such as gradient descent to minimize a loss function that measures the difference between the predicted outputs and the true labels of the training examples. The network is then evaluated on a separate validation set to tune the hyperparameters and prevent overfitting. MLPs are powerful and flexible models that can capture complex patterns in high-dimensional data, but they can also be computationally expensive and require careful tuning of the architecture and hyperparameters to achieve good performance [1, 14, 15]. The learning mechanism utilized by the multilayer perceptron is recognized as the "generalized delta rule" or the "backpropagation rule". This rule iteratively computes an error metric for each input and propagates this error from one layer to the preceding layer. The adjustments made to the weights of a specific node are directly proportional to the error observed in the units to which it is linked.

Let:

$E_p$ = error function for pattern $p$
$t_{pj}$ = target output for pattern $p$ on node $j$
$o_{pj}$ = actual output for pattern $p$ on node $j$
$w_{ij}$ = weight from node $i$ to node $j$

The error function $E_p$ is defined to be proportional to the square of the difference $t_{pj}$ - $o_{pj}$

$$E_p = \frac{1}{2}\Sigma(t_{pj} - o_{pj})^2 \qquad (4)$$

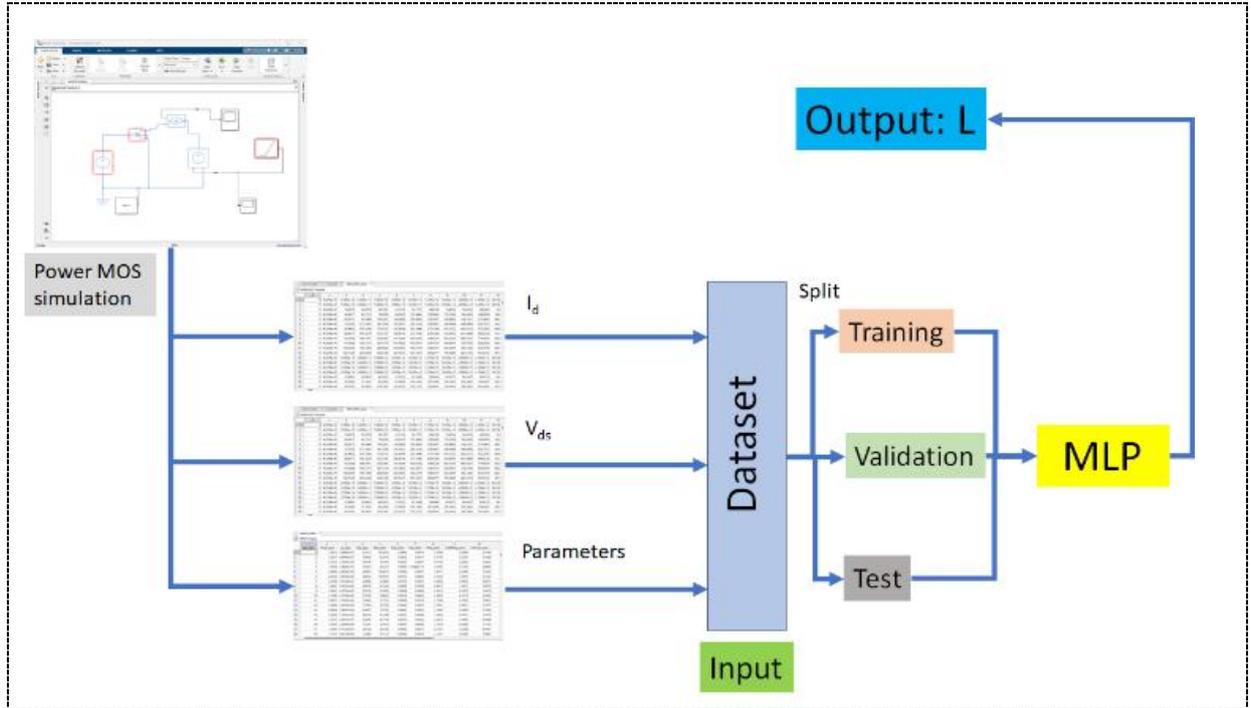
**Figure 3.** Input and output for the implemented MLP model.

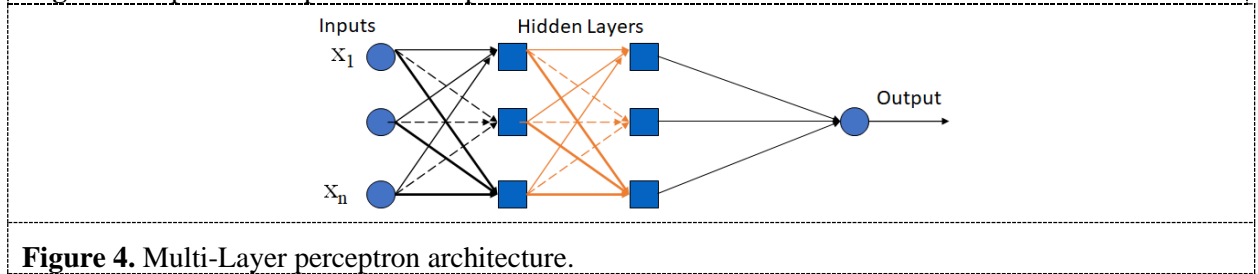
**Figure 4.** Multi-Layer perceptron architecture.

The activation of each unit $j$, for pattern $p$, can be written as:

$$net_{pj} = \sum(w_{ij} - o_{pi}) \qquad (5)$$

The output from each unit $j$ is determined by the non-linear transfer function $f_j$:

$$o_{pj} = f_j(net_{pj}) \qquad (6)$$

We assume $f_j$ to be the sigmoid function,

$$f(net) = \frac{1}{(1+e^{-k.net})} \qquad (7)$$

Here, $k$ denotes a constructive constant regulating the function's "spread". The delta rule enacts weight adjustments that trace the trajectory of the most rapid decrease on a weight space surface. The elevation at any given location on this surface corresponds to the error metric $E_p$. This correspondence can be demonstrated by revealing that the derivative of the error metric concerning each weight corresponds to the weight adjustment determined by the delta rule, involving a negative proportionality constant, as expressed:

$$\Delta_p w_i \propto -\frac{\partial E_p}{\partial w_{ij}} \quad (8)$$

For the proposed architecture MLP model (Figure 4) has been applied with the setup involving different hyperparameters (see TABLE 2):

Table 2. MLP Model setup

| MLP Model parameter | Values |
| --- | --- |
| Multi Layer Perceptron layers | 6 |
| Hidden layers | 4 |
| Batch size | 32 |
| Input size | 100 |
| Learning rate | 0.0001 |
| Optimizer | Adam |
| Epochs | 100 |

It has been chosen to employ ReLU as the activation function for the MLP.

## 4. Experimental results

We conduct extensive experiments based on the prepared dataset. To demonstrate the accuracy of the MLP model, we compare it in terms of the MAE performance metrics with another flow predictors, which is the Temporal Convolutional Network (TCN). Note that the TCN model achieves the state-of-the-art forecasting result [11]. All methods have been implemented on a Python environment on a PC with Intel(R) Core(TM) i7 CPU, 16 GB memory and NVIDIA RTX 2050 GPU as in [8, 9]. We build the power MOSFET model with Matlab Simulink, showed in Figure 3. The simulated model returns $I_d$-$V_{ds}$ curves without a standardized methodology (i.e., without standardize $V_{ds}$ to $I_d$), ending up with curves of different lengths. The dataset has been created by taking a constant temperature of 25° and varying all the parameters listed below at different Gate Voltage $V_{gs}$ (from 1 to 12 Volts), obtaining 12 curves for each device. In Figure 2 it is showed the simulated $I_d$ data plotted on curves. The dataset has been split in training (80% of the total dataset), validation (10%) and test (10%) subsets and processed during training and evaluation of the MLP model. Finally, MLP return one specific Power MOS parameter: $L_p$.

*4.1. Define the input parameters*

The input parameters for a MOSFET are typically defined by the drain-source voltage ($V_{ds}$) and the gate-source voltage ($V_{gs}$) (Figure 3). The drain current ($I_d$) is then determined by the current flowing through the device from the drain to the source terminals, which is controlled by the gate-source voltage. The $V_{ds}$ and $V_{gs}$ are typically specified by the manufacturer and can be found in the device datasheet. To define the input parameters, it has been set on Simulink, the appropriate voltage levels for $V_{ds}$ and $V_{gs}$ and then after model simulation collected the resulting drain current $I_d$.

*4.2. Define the output parameters*

The output parameters for a MOSFET to solve for, are the transconductance ($KP$), the drain-source resistance ($R_d$, $R_s$), the channel length ($L_p$), the channel width $W_p$, the voltage threshold $V_t$, the mobility degradation factor $\theta$, the Surface inversion potential $\phi$, the body effect factor $\gamma$ (Figure 3).

*4.3. Use the MOSFET's device equations*

Simulink simulations are slow (at least 1 second is needed for a single simulation), by studying SPICE Level 3 equations we can implement them in a Python environment by varying the set of parameters values. Also, the returned curves can be standardized by the $V_{ds}$, avoiding the aforementioned problem in Simulink, by choosing $V_{ds}$ granularity (varying by 1 Volt or 0,1 Volt or 0,01 Volt and so on), to express the output parameters in terms of the input parameters. These equations will typically include the channel length modulation effect and the saturation current as in (1).

*4.4. MLP model implemented pipeline*

The implemented pipeline for the MLP model includes the following steps (Figure 1):
*Data Preparation*: The first step is to prepare the data for training the MLP. This involves tasks such as data cleaning, feature engineering, and splitting the data into training and testing sets.
*Build model architecture*: The next step is to define the architecture of the MLP. This involves specifying the number of layers, the number of neurons in each layer, the activation functions to be used, and the type of regularization to be applied.
*Initialization*: The weights and biases of the MLP need to be initialized before training begins.
*Forward propagation*: During training, the MLP takes in inputs and propagates them forward through the layers of neurons. The output of each layer is computed using the weights and biases of that layer and passed on to the next layer.
*Backward propagation*: After forward propagation, the MLP calculates the error between the predicted output and the actual output. This error is then propagated backwards through the layers of neurons to update the weights and biases using gradient descent.
*Optimization*: During training, various optimization techniques can be used to improve the performance of the MLP. These include techniques such as stochastic gradient descent, batch gradient descent, and adaptive learning rate methods like Adam.
*Evaluation*: After training, the performance of the MLP is evaluated on a separate test set to determine its accuracy and generalization ability.
*Hyperparameter tuning*: Finally, hyperparameters such as the learning rate, batch size, and regularization strength can be tuned to further improve the performance of the MLP. This can be done using techniques such as grid search or random search.
In calculating the error of the model during the optimization process, a loss function must be chosen. We choose MSE loss function and in the training, validation and test phase, has been measured the Mean Squared Loss error, through a criterion that measures the mean squared error (squared L2 norm) between each element in the input x and target y.
After 100 epochs, results showing that training, validation and test MSE loss converge. The plots in Fig. 5 make the situation clearer. It looks as though the line plot for the training set is dropping to converge with the line for the validation and test set. It means that prediction and target converge with a minimum loss error.

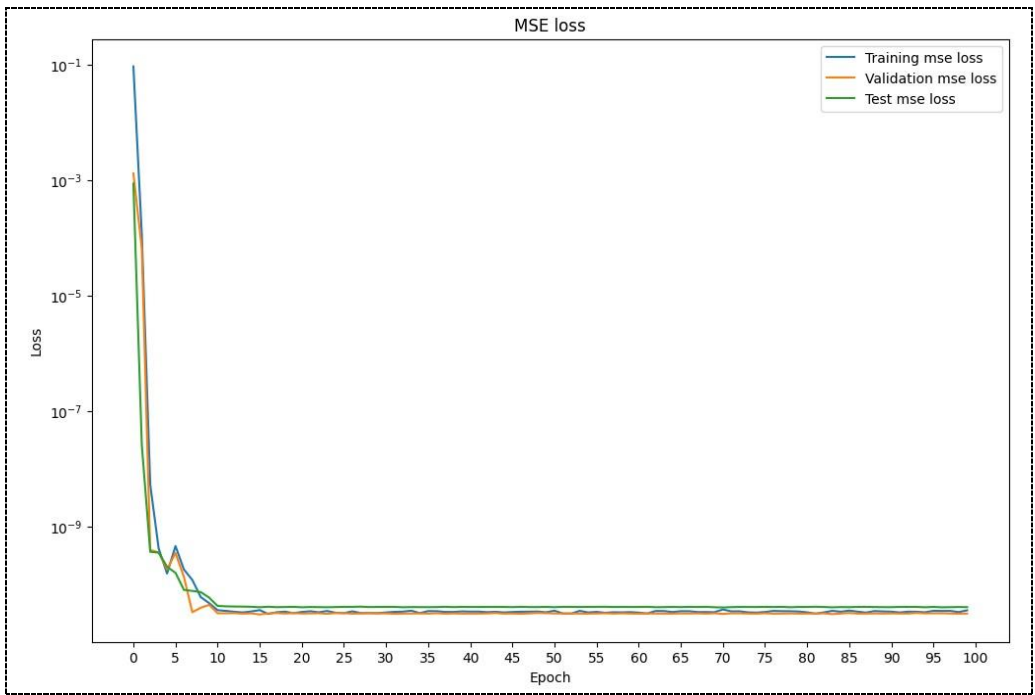

**Figure 5.** MSE Loss for training, validation and test for 100 epochs

In Figure 6, and Figure 7 it is reported plot of MSLE (Mean Squared Logarithmic Error), and MAE (Mean Absolute Error) which shows loss convergence to $1*10*e-8$ for training validation and test.

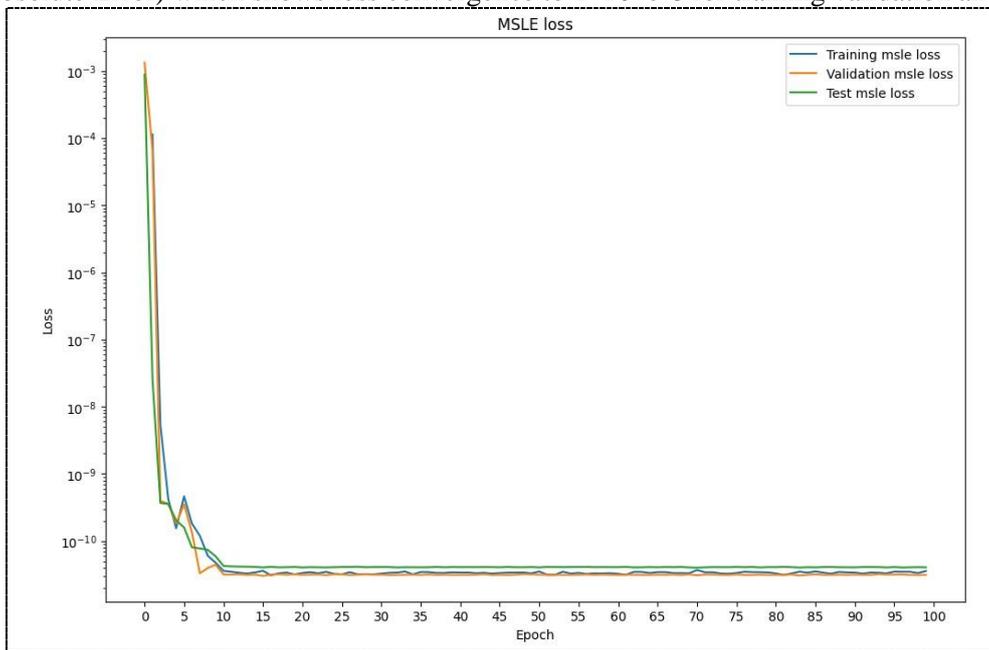

**Figure 6.** MSLE Loss for training, validation and test for 100 epochs

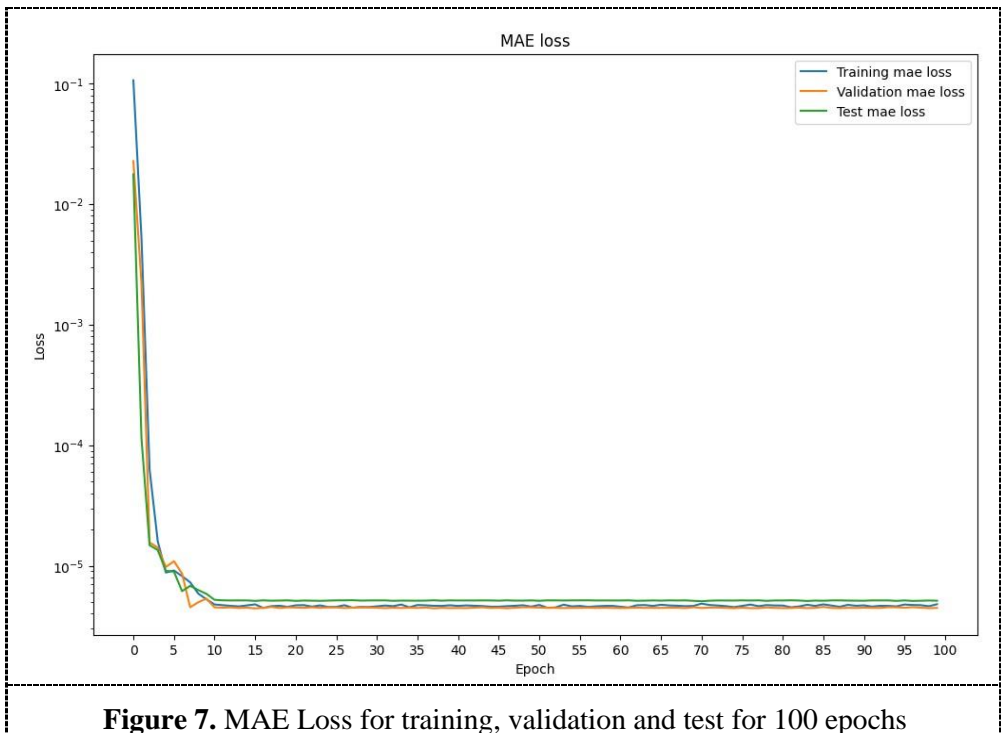

**Figure 7.** MAE Loss for training, validation and test for 100 epochs

In Figure 8 it is reported plot of the MAPE (Medium Absolute Percentage Error) which shows a convergence of the percentage error closed to 8% for training validation and test.

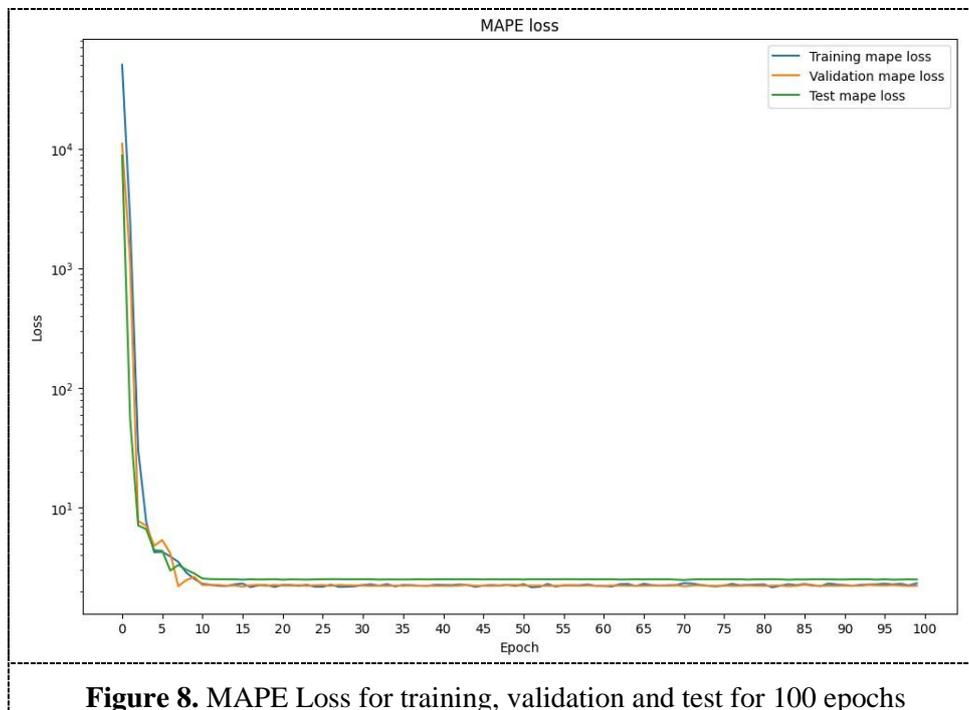

**Figure 8.** MAPE Loss for training, validation and test for 100 epochs

## 5. Conclusions and future works

Based on the good results obtained from the inverse modelling of SiC power MOS with a multi-layer perceptron (MLP) to determine the length of a physical device, the following conclusions can be drawn: MLP is a powerful tool for inverse modelling of SiC power MOS devices. It can accurately predict the length of a physical device from electrical measurements, which can greatly simplify the design process. The performance of the MLP model is highly dependent on the quality of the training data. Therefore, careful selection and pre-processing of the data are critical for achieving good results. The use of MLP for inverse modelling of SiC power MOS can significantly reduce the time and cost associated with device fabrication and testing. The MLP model can be extended to other types of semiconductor devices for similar inverse modelling tasks.

In terms of future work could be explored area related to the development of more advanced MLP models that can capture the complex relationships between device parameters and electrical measurements. Could be investigated the impact of different types of noise on the performance of the MLP model, and the development of techniques to mitigate their effects. Could be explored other machine learning algorithms that can be used for inverse modelling of semiconductor devices, such as convolutional neural networks or recurrent neural networks. Finally, could be extended the MLP model to handle multi-dimensional input data, such as images or spectroscopic measurements, which can provide more comprehensive information about the device under test.

Future works aims to replace the proposed deep backbone with more performer deep pipeline which embeds self-attention mechanisms [17,18].

In Table 3 it is reported the best accuracy results comparing a Temporal Convolutional Network (TCN), a Long Short Term (LSTM) and a Multi-Layer Perceptron (MLP) model, the MLP achieved the best accuracy with a lower mean squared error (MSE) compared to the TCN and LSTM. Despite the TCN's ability to capture long-term temporal dependencies, the MLP's flexible architecture and nonlinear activation functions allowed it to outperform the TCN in this particular experiment.

**Table 3.** Performance benchmarks

| Model | Accuracy |
|---|---|
| TCN | $5.811e^{-10}$ |
| LSTM | $6.631e^{-8}$ |
| MLP | $1.581e^{-11}$ |